\documentclass[onecolumn,prb,showpacs]{revtex4}
\usepackage{graphicx,psfrag}
\usepackage{slashed,amssymb,amsmath,dsfont}
\def\no{\noindent}
\newcommand{\nn}{\nonumber}

\begin{document}

\title{
Finite-size scaling in a 2D disordered electron gas with spectral nodes
}

\author{Andreas Sinner and Klaus Ziegler}
\affiliation{Institut f\"ur Physik, Universit\"at Augsburg\\D-86135 Augsburg, Germany 
}

\date{\today}

\begin{abstract}
We study the DC conductivity of a weakly disordered 2D electron gas with two bands  and  spectral nodes,
employing the field theoretical version of the Kubo--Greenwood conductivity formula. Disorder
scattering is treated within the standard perturbation theory by summing
up ladder and maximally crossed diagrams. The emergent gapless (diffusion) modes determine
the behavior of the conductivity on large scales. We find a finite conductivity with an intermediate
logarithmic finite-size scaling towards smaller conductivities but do not obtain
the logarithmically divergence of the weak-localization approach. Our results agree with
the experimentally observed logarithmic scaling of the conductivity in graphene with the formation
of a plateau near $e^2/\pi h$. 
\end{abstract}


\maketitle

\section{introduction}

Transport in two-dimensional (2D) electronic systems has been a subject of intense research for several decades.
One of the reasons for the attractiveness of this field is that quantum interference is strong in 2D and 
interesting phenomena exist, such as the quantum Hall effect. Despite of its long history, some aspects of
electronic transport are still puzzling. 
For some time there was a consensus about the role disorder plays in transport processes, dominated
by Anderson localization of electronic wave functions for conventional 2D systems~[\onlinecite{Abrahams1979,Gorkov1979,Hikami1980}].
A first hint, though, for an unconventional behavior was the transition between Hall plateaux in quantum Hall systems, which
indicated the existence of a metallic state in a 2D electronic system under special conditions~[\onlinecite{Hanein1998}].
Even more important was the discovery of metallic states in graphene [\onlinecite{Geim2005,kim07,elias09}] and in a number of 
chemical compounds, commonly referred to as topological insulators~[\onlinecite{Allen2010,Chen2012,Hasan2010,Qi2011}],
where the band structure has nodes and the dispersion of electronic quasiparticles is linear in the vicinity of these nodes.
Although these compounds represent pristine 2D systems, they reveal a finite DC conductivity which is very robust against
thermal fluctuations and disorder. In the course of subsequent years these systems underwent careful studies from both 
the experimental and the theoretical point of view which clearly indicate that the finite DC value is a robust property. 
In the experiments the following features have been observed for the conductivity:
1) It decreases with increasing sample size to the DC value, starting from some value considerably higher than
this~[\onlinecite{Heer2014}]; 
2) For a finite sample it exhibits a pronounced temperature dependence, decreasing logarithmically with decreasing temperature
to a plateau at low temperatures~[\onlinecite{Savchenko2010,Fuhrer2009,Fuhrer2011,Camassel2011,Liu2014,Novoselov2012,Novoselov2011,Kogan2015}].
The plateau value varies only slightly from sample to sample. However, the latter effect is also observable in conventional 
2D metals where this saturation is usually linked to the presence of magnetic impurities~[\onlinecite{Gantmakher}]. 
The same argumentation is sometimes used in the context with the thermal conductivity saturation in 2D electron gases 
with nodal points~[\onlinecite{Kogan2015}]. 
This argument gives rise to the question of what would happen if we remove these impurities: Would the conductivity decrease
to negative values logarithmically without limit?
In the field--theoretical approach it is natural to use the density--density Kubo formula,
which predicts a nonzero plateau~[\onlinecite{Wegner1979a,McKane1981,Fradkin1986,Ludwig1994,Ziegler1997}], whose values are always non-negative.
The current--current Kubo formula, on the other hand, leads to an infinite negative conductivity in the weak--localization approach. 
This unphysical result is usually avoided by a phenomenological inelastic scattering cut-off for small momentum transfer
[\onlinecite{Gorkov1979,Altshuler1981,Hikami1980,Altshuler1994,Efetov1997,Levitov,Shon1998,Ando2002,McCann2006,Khveshchenko2006,Tkachov2011,Schmeltzer2013}].
Here it should be noted that the discrepancy between the Kubo formulae was mentioned previously in Refs.~[\onlinecite{Ludwig1994,PhysRevB84,PhysRevB89}], 
and its origin is that disorder averaging and the DC limit do not commute.
The density--density Kubo formula has also the advantage that it can be derived directly from diffusion. This makes it more appropriate for the analysis of the DC transport properties. 

\vspace{1mm}
\no
In the following we will pursue and develop further the idea of the perturbative weak scattering approach to transport in 2D disordered Dirac 
electron gases and its scaling behavior, first formulated in our recent papers~[\onlinecite{PhysRevB81,PhysicaE71,PhysRevB90}]. 
This approach follows closely the non--linear sigma model
concept~[\onlinecite{Wegner1979a,Wegner1979,Wegner1980,McKane1981,Hikami1981,Fradkin1986,Ziegler1997,Ziegler2009,Ziegler2012,PhysRevB86}],
leading in the case of 2D Dirac fermions to the same expressions for the massless modes as the perturbative disorder averaging 
technique~[\onlinecite{Gorkov1979,Altshuler1981,Hikami1980,Altshuler1994,Efetov1997,Levitov,Shon1998,Ando2002,McCann2006,Khveshchenko2006,Tkachov2011,Schmeltzer2013}].
The difference in comparison with weak--localization approach is the implementation of the density--density Kubo formula instead of the
current--current Kubo formula. The finite--size scaling behavior of the conductivity is then recovered via the coarse graining procedure from expressions obtained in continuous limit and reproduces astonishingly accurately the experimentally observed conductivity behavior, not only qualitatively but quantitatively as well.

\vspace{1mm}
\no
In general, on bipartite lattices Dirac cones appear pairwise. In order for the Hamiltonian to remain invariant under the time--reversal transformation both Dirac cones must be linked to each other by a parity transformation, i.e. they have different chiralities.
Then all possible scattering processes can be roughly subdivided into those with and without mixing of quasiparticles with different chiralities. Previous studies are inconclusive on the role such processes may play in electronic transport, ranging from essentially no difference~[\onlinecite{Shon1998}] to strong statements about a localization--antilocalization transition~[\onlinecite{McCann2006,Ando2002,Khveshchenko2006}]. 
General symmetry arguments indicate that massless modes should survive in the presence of random internode scattering (different chiralities)~[\onlinecite{PhysicaE71}]. We will address this point in more details elsewhere.

\vspace{1mm}
\no
In the case of intra--node scattering (same chirality), which is the main object of investigation in this paper, we observe the total suppression of the conductivity contributions arising from the backward scattering channel, which is associated with
diagrams with maximally crossed impurity lines~[\onlinecite{Levitov}]. Consequently, the conductivity is uniquely 
determined by the contributions from another channel which is expressed in terms of ladder diagrams, 
and for both considered cases they are equal. However, for other disorder types the role of both channels is different.

\vspace{1mm}
\no
The structure of the paper is as follows: In Sections~\ref{sec:Kubo} and \ref{sec:WSA} we define the model and perform the general evaluation of the Kubo--Greenwood conductivity formula irrespective of the disorder type. 
In Section~\ref{sec:Example} we discuss the conductivity for intra--node scattering potentials. 
In Section~\ref{sec:Scaling} we extract by the coarse graining procedure the conductivity scaling functions from the expressions obtained in the continuous limit.

\section{The Kubo--Greenwood conductivity formula}
\label{sec:Kubo}

Motivated by diffusion, we consider the field theoretical version of the Kubo--Greenwood conductivity 
formula~[\onlinecite{Wegner1979a,McKane1981,Fradkin1986,Ludwig1994,Ziegler1997,PhysicaE71}]
\begin{equation}
\label{eq:KuboGen1} 
\bar\sigma^{}_{\mu\mu}(\omega) = \frac{e^2}{h}\int_{}^{}dE~\Gamma^{}_\mu(E,\omega)~\frac{f(E)-f(E+\omega)}{\omega},
\end{equation}
where $\omega$ denotes the physical frequency, $f(E)$ the Fermi function, and the disorder averaged kernel
\begin{equation}
\label{eq:KuboGen2}
\Gamma^{}_\mu(E,\omega) = -\omega^2{\rm Tr}\sum_r~r^2_\mu~\langle G^+(r,0)G^-(0,r)\rangle,
\end{equation}
with advanced (retarded)  single--particle Green's function $G^+(r,r^\prime)$ ($G^-(r,r^\prime)$) 
\begin{equation}
G^{\pm}(r,r^\prime) = \langle r| [E \pm i\epsilon + H + V ]^{-1}| r^\prime\rangle, 
\label{green_func}
\end{equation}
where $i\epsilon = \omega+i0^{+}$ denotes the zero--temperature Matsubara frequency, and $H$ is the low--energy approximation of a tight--binding Hamiltonian 
\begin{equation}
\label{eq:TBH} 
H^{}_{} = -t\sum_{\langle rr^\prime\rangle}~(c^\dag_r d^{}_{r^\prime} + d^\dag_{r^\prime} c^{}_{r} )
\ .
\end{equation}
$c$ and $d$ ($c^\dagger$ and $d^\dagger$) are the fermionic annihilation (creation) operators with respect to each sublattice of a bipartite lattice, respectively.
The neighboring lattice sites $r$ and $r^\prime$ are connected with the hopping amplitude $t$, and the summation is performed over nearest neighbor pairs only. 
Finally, $V$ denotes the one particle random potential. Due to the randomness each realization of the system lacks translational invariance and the Green's function depends on both sites $r$ and $r'$. We assume a Gaussian distribution independently for each site with
\begin{equation}
\label{eq:Correlator}
\langle V^{}_{ab}(r)\rangle = 0,\hspace{5mm}  
\langle V^{}_{ab}(r) V^{}_{a^\prime b^\prime}(r^\prime) \rangle = g\delta(r-r^\prime) \Sigma^{}_{ab}\Sigma^{}_{a^\prime b^\prime},
\end{equation}
where $g$ is referred to as the disorder strength and $\Sigma^{}$ represent matrices on the extended spinor space. 
In this work we restrict the consideration to the case of the honeycomb lattice with the Fermi energy laying at the nodal points.
This yields the effective low--energy Hamiltonian~[\onlinecite{Wallace1947,Semenoff1984}] 
\begin{equation}
H = 
\hbar\left( 
\begin{array}{ccc}
p^{}_1\sigma^{}_1+p^{}_2\sigma^{}_2 & & 0 \\
\\
0 & &  p^{}_1\sigma^{}_1-p^{}_2\sigma^{}_2
\end{array}
\right),
\end{equation}
where the momentum operators are $p^{}_i = -iv_F\nabla^{}_i$, $v_F$ denoting the Fermi velocity, and $\sigma^{}_i$ are Pauli matrices. 
Below we use dimensionless energy units, i.e. $\hbar v^{}_F=1$ in units of inverse length. Then the disorder strength is 
measured in units of $(\hbar v^{}_F)^2$, where the quantity $s=(\hbar v^{}_F)^2$ represents an appropriate reference scale. The randomness is supposed to be the local fluctuating chemical potential with $\Sigma=\mathds{1}$.

\section{Weak random scattering approach}
\label{sec:WSA}

We approach the DC conductivity within a weak scattering approach, in which the disorder average is performed perturbatively. 
At frequencies small as compared to the typical band width of the clean system and below room temperature we can employ the usual 
approximation $f(E+\omega) \sim f(E) +\omega\delta(E)$, which trivializes the energy integral in Eq.~(\ref{eq:KuboGen1}). 
Then the conductivity formula becomes
\begin{eqnarray}
\label{eq:Kubo1}
\bar\sigma^{}_{\mu\mu} &=& \frac{e^2}{h}\lim_{\epsilon\to0}\epsilon^2\left.\left(-\frac{\partial^2}{\partial q^2_\mu}\right)\right|_{q=0} \sum_r e^{iq\cdot r} 
\langle G^+_{nj}(r,0) G^-_{jn}(0,r)\rangle,
\end{eqnarray}
where we have used the Fourier representation of the position operator and the summation convention for matrix elements with respect to the spinor index. The averaged two-particle Green's function $\langle G^+_{nj}(r,0) G^-_{jn}(0,r)\rangle$ can be treated within a perturbation expansion
in powers of a weak scattering rate $\eta$. Using a summation of ladder and maximally crossed diagrams lead to 
\begin{eqnarray}
\nn
\bar\sigma^{}_{\mu\mu} &=& \frac{e^2}{h}\lim_{\epsilon\to0}{\epsilon^2}\left.\left(-\frac{\partial^2}{\partial q^2_\mu}\right)\right|_{q=0}\sum_{rr^\prime}e^{iq\cdot r}~\bar G^+_{ij}(r^\prime,0)\bar G^-_{kn}(0,r^\prime)\\
\label{eq:ATPGF2}
&\times&\left[\left(\mathds{1}-g[\bar G^+\Sigma][\bar G^-\Sigma]\right)^{-1}_{rr^\prime|nj;ik} 
+ \left(\mathds{1}-g[\bar G^+\Sigma][\bar G^{-}\Sigma]^{\rm {\bf T}}\right)^{-1}_{rr^\prime|nk;ij}\right],
\end{eqnarray}
where the full transposition operator $\rm\bf T$ applies to all degrees of freedom, i.e. to the spatial ones as well. The renormalized one--particle Green's functions read
\begin{equation}
\bar G^{\pm}(r,r^\prime) = \langle r| [\pm iz + H ]^{-1}| r^\prime\rangle ,
\end{equation}
with $z=\epsilon+\eta$. The scattering rate $\eta$ is determined self--consistently from 
\begin{equation}
\pm i\frac{\eta}{g} = - \langle r| [\pm iz + H ]^{-1}| r\rangle = \int\frac{d^2p}{(2\pi)^2}~\frac{\pm iz}{p^2+z^2}.
\end{equation}
In the field theoretic language this condition defines the saddle point of the functional integral, around which an expansion in powers of a small $\eta$ 
can be performed. For $\epsilon\sim0$ we get
\begin{equation}
\label{eq:SCBA2}
1 - g\int\frac{d^2p}{(2\pi)^2}~\frac{1}{p^2+z^2} \sim \frac{\epsilon}{\eta}.
\end{equation}
We introduce for the terms in the second line of Eq.~(\ref{eq:ATPGF2})
\begin{eqnarray}
\label{eq:LC}
a)\hspace{2mm}
t=g[\bar G^+\Sigma][\bar G^-\Sigma] \ , \ \ \ 
b)\hspace{2mm}
\tau=g[\bar G^+\Sigma][\bar G^{-}\Sigma]^{\rm {\bf T}}
\end{eqnarray}
the ladder Eq.~(\ref{eq:LC}a) and the maximally crossed Eq.~(\ref{eq:LC}b) channel matrices, respectively~[\onlinecite{PhysicaE71}].
The matrices $t$ and $\tau$ read in terms of their Fourier components:
\begin{eqnarray}
t^{}_{r^\prime r|ab;cd} &=& g\int\frac{d^2q}{(2\pi)^2}~e^{-iq\cdot(r-r^\prime)}
\int\frac{d^2p}{(2\pi)^2}~[\bar G^+(p)\Sigma]^{}_{ac}[\bar G^{-}(q+p)\Sigma]^{}_{bd}
\ , \\
\tau^{}_{r^\prime r|ab;cd} &=& g\int\frac{d^2q}{(2\pi)^2}~e^{-iq\cdot(r^\prime-r)}
\int\frac{d^2p}{(2\pi)^2}~[\bar G^+(p)\Sigma]^{}_{ac}[\bar G^{-}(q-p)\Sigma]_{db}
\ ,
\label{weak_scatt0}
\end{eqnarray}
The different signs in the argument of $t$ and $\tau$ are a consequence of the transposition on the position space in the MC--channel. 
The corresponding Fourier transformed matrices $\mathds{1}-{\tilde t}_{q}$ and $\mathds{1}-{\tilde \tau}_{q}$ for $\epsilon=0$ and $q=0$ read
\begin{eqnarray}
\label{eq:MLC}
M^{LC}_{ab;cd} &=& \left. {\delta^{}_{ac}\delta^{}_{bd}} - {g}\int\frac{d^2p}{(2\pi)^2}~[\bar G^+(p)\Sigma]^{}_{ac}[\bar G^{-}(p)\Sigma]^{}_{bd} \right|_{\epsilon=0},\\
\label{eq:MMC}
M^{MC}_{ab;cd} &=& \left. {\delta^{}_{ac}\delta^{}_{bd}} - {g}\int\frac{d^2p}{(2\pi)^2}~[\bar G^+(p)\Sigma]^{}_{ac}
[\bar G^{-}(-p)\Sigma]^{}_{db} \right|_{\epsilon=0}
\ .
\end{eqnarray}
The eigenvalues of matrices $M$ provide a decay length for the full matrices in Eq.~(\ref{eq:LC}). In particular, 
a vanishing (e.g. gapless) eigenvalue gives a long-range diffusion-like behavior, which gives
a non-vanishing contribution to the conductivity. Massive modes, on the other hand, do not 
contribute to the conductivity the limit $\epsilon\to 0$ due to the prefactor $\epsilon^2$ in 
Eq. (\ref{eq:Kubo1}). Therefore, massive modes can be neglected subsequently.
Depending on the type of disorder there may or may not be gapless modes. 
If gapless modes exist, then for small momenta and frequencies we get in the ladder--channel in diagonal representation (in the 
channel of maximally crossed diagrams analogously) 
\begin{equation}
\mathds{1}_N-{\tilde t}_q\sim \left(\frac{\epsilon}{\eta} + gD^{}_0q^2\right) \mathds{1}^{}_N
\end{equation}
for $N$ massless channels. $D^{}_0$ is the expansion coefficient:
\begin{equation}
\label{eq:DiffCoef}
D^{}_0 = \frac{1}{2} \int\frac{d^2p}{(2\pi)^2}~\frac{1}{[p^2+\eta^2]^2}
\ .
\end{equation}
Thus, the conductivity becomes
\begin{equation}
\bar\sigma^{}_{\mu\mu} = \frac{e^2}{h} \int\frac{d^2p}{(2\pi)^2}~\bar G^+_{ij}(p)\bar G^-_{kn}(p)
\lim_{\epsilon\to0}\epsilon^2\left.\left(-\frac{\partial^2}{\partial q^2_\mu}\right)
\left[(\mathds{1}-{\tilde t})^{-1}_{q|nj,ik}+(\mathds{1}-{\tilde \tau})^{-1}_{q|nk,ij}\right]
\right|_{q=0}
\ .
\label{eq:gen_cond0}
\end{equation}

\section{Conductivity for particular disorder types} 
\label{sec:Example}

\no
Below we investigate the effect of the intra-node scattering. 
For this discussion it is useful to introduce the matrix notation
\begin{equation}
A=\begin{pmatrix}
A_{11,11} & A_{11,12} & A_{11,21} & A_{11,22} \cr
A_{12,11} & A_{12,12} & A_{12,21} & A_{12,22} \cr
A_{21,11} & A_{21,12} & A_{21,21} & A_{21,22} \cr
A_{22,11} & A_{22,12} & A_{22,21} & A_{22,22} \cr
\end{pmatrix}
\ .
\end{equation}
In the simplest case, with diagonal disorder and in the absence of the inter--node scattering the calculation reduces to 
a single cone only. Then the Dirac propagators $\bar G^\pm(p)$ read in Fourier representation
\begin{equation}
\bar G^\pm(p) = \frac{1}{p^2+z^2} \left(
\begin{array}{ccc}
 \mp iz & & p^{}_1 - i p^{}_2 \\
 \\
 p^{}_1 + i p^{}_2 & & \mp iz
\end{array}
\right),
\end{equation}
where $z=\epsilon+\eta$. The limit $\epsilon\to0$ yields for the mass matrices in both channels:
\begin{eqnarray}
\label{eq:MMLC}
M^{LC} = 
\left( 
\begin{array}{cccc}
\alpha& 0   & 0  & 0  \\
0   & \alpha  & -\alpha & 0 \\
0   & -\alpha  & \alpha &  0 \\
0   & 0 & 0 & \alpha
\end{array}
\right)^{}, &\;\; &
M^{MC} = 
\left(
\begin{array}{cccc}
\alpha & 0 & 0 & \alpha \\
0 & \alpha & 0 & 0 \\
0 & 0 & \alpha & 0  \\
\alpha & 0 & 0 & \alpha
\end{array}
\right)^{},
\;\;\; \alpha =g\int\frac{d^2p}{(2\pi)^2}~\frac{p^2}{[p^2+\eta^2]^2},
\end{eqnarray}
i.e. each of them has a single zero eigenvalue. Moreover, for $q\sim 0$ we get 
\begin{equation}
\label{eq:tq}
\mathds{1}-{\tilde t}_q\sim  \left( 
\begin{array}{cccc}
\displaystyle \alpha + \frac{\epsilon}{\eta} + D^\prime q^2 & i\gamma q e^{-i\varphi^{}_q} & i\gamma qe^{-i\varphi^{}_q} & 
\zeta q^2e^{-i2\varphi^{}_q} \\
i\gamma q e^{i\varphi^{}_q}&\displaystyle \alpha + \frac{\epsilon}{\eta} + D^\prime q^2 &-\alpha+D^{\prime\prime}q^2 & i\gamma q e^{-i\varphi^{}_q} \\
i\gamma qe^{i\varphi^{}_q}  & -\alpha+D^{\prime\prime}q^2 & \displaystyle \alpha + \frac{\epsilon}{\eta} + D^\prime q^2 & i\gamma qe^{-i\varphi^{}_q}  \\
\zeta q^2e^{i2\varphi^{}_q} & i\gamma qe^{i\varphi^{}_q}  & i\gamma q e^{i\varphi^{}_q} & \displaystyle \alpha + \frac{\epsilon}{\eta} + D^\prime q^2 
\end{array}
\right),
\end{equation}
and 
\begin{equation}
\mathds{1}-{\tilde\tau}_q \sim \left( 
\begin{array}{cccc}
\displaystyle \alpha + \frac{\epsilon}{\eta} + D^\prime q^2 & i\gamma q e^{i\varphi^{}_q} & -i\gamma qe^{-i\varphi^{}_q} & \alpha-D^{\prime\prime}q^2 \\
i\gamma q e^{-i\varphi^{}_q}&\displaystyle \alpha + \frac{\epsilon}{\eta} + D^\prime q^2 & -\zeta q^2e^{-i2\varphi^{}_q}& -i\gamma q e^{-i\varphi^{}_q} \\
 -i\gamma qe^{i\varphi^{}_q}  & -\zeta q^2e^{i2\varphi^{}_q}   & \displaystyle \alpha + \frac{\epsilon}{\eta} + D^\prime q^2 & i\gamma qe^{i\varphi^{}_q}  \\
 \alpha-D^{\prime\prime}q^2 & -i\gamma qe^{i\varphi^{}_q}  & i\gamma q e^{-i\varphi^{}_q} & \displaystyle \alpha + \frac{\epsilon}{\eta} + D^\prime q^2 
\end{array}
\right),
\end{equation}
where
\begin{eqnarray}
\nn
&\displaystyle
D^\prime = \frac{g}{6}\int\frac{d^2p}{(2\pi)^2}~\frac{2\eta^2}{[p^2+\eta^2]^3},\;\;\; D^{\prime\prime} = \frac{g}{6} \int\frac{d^2p}{(2\pi)^2}~\frac{\eta^2+3p^2}{[p^2+\eta^2]^3}, &\\
\nn
&\displaystyle
\varphi = {\rm atan}\left[\frac{q^{}_y}{q^{}_x}\right],\;\;\; 
\gamma = \frac{g}{2}\int\frac{d^2p}{(2\pi)^2}~\frac{\eta}{[p^2+\eta^2]^2}, \;\;\; 
\zeta = \frac{g}{6}\int\frac{d^2p}{(2\pi)^2}~\frac{1}{[p^2+\eta^2]^2}.&
\end{eqnarray}
Inserting these expressions into our general conductivity expression (\ref{eq:gen_cond0}) gives eventually
\begin{eqnarray}
\label{eq:LCres}
\bar\sigma^{LC}_{\mu\mu} &\sim& 4\eta^2 D^{}_0 \frac{e^2}{h}, \\
\label{eq:MCres}
\bar\sigma^{MC}_{\mu\mu} &\sim& 0.
\end{eqnarray}
In Eq.~(\ref{eq:LCres}) terms of the order $1/\alpha,\;\;\alpha\sim\log p$, which arise from the linear order of gradient expansion in Eq.~(\ref{eq:tq}) are neglected due to their smallness. Eq.~(\ref{eq:MCres}) reveals the total suppression of the conductivity contribution (except for possible higher order contributions) from the backward scattering processes, with which maximally crossed diagrams are usually
associated~[\onlinecite{Altshuler1994,Efetov1997}].

\vspace{1mm}
\no
The most striking feature of the obtained results is the absence of any logarithmic divergences in the conductivities.
However, the question remains how the logarithmic behavior, observed in a number of experiments for positive values of the conductivity, can be explained within our approach? In the next section this will be described within a finite-size scaling approach, following closely the argumentation of our previous work Refs.~[\onlinecite{PhysRevB81,PhysRevB90}].

\section{Scaling behavior of transport properties}
\label{sec:Scaling}

\no
The evaluation of Eq.~(\ref{eq:LCres}) for the infinite-size limit gives for the DC conductivity $\bar\sigma^{}_0$, which was predicted 
theoretically~ [\onlinecite{Wegner1979a,Fradkin1986,Ludwig1994}] and measured experimentally~[\onlinecite{kim07,elias09}].
In order to compare with experimental measurements we must also consider finite-size effects.
In graphene, for instance, a typical sample size is of the order of several micrometers. 
To include the finite size of the sample in our theory, we must introduce a discretization of the momentum integral in the diffusion coefficient 
for a square of size $2L\times2L$:
\begin{equation}
\label{eq:DiffDiscret}
D^{}_0 = \frac{1}{2}\int\frac{d^2p}{(2\pi)^2}~\frac{1}{[p^2+\eta^2]^2} \to 
\frac{1}{2}\frac{1}{(2L)^2} \sum^{L}_{n,m=-L}~\frac{1}{[k^2_n + k^2_m + \eta^2]^2},
\end{equation}
with the scattering rate $\eta$ evaluated from the saddle--point equation
\begin{equation}
\label{eq:SPDiscret}
\frac{1}{g} = \int\frac{d^2p}{(2\pi)^2}~\frac{1}{p^2+\eta^2}\to 
\frac{1}{(2L)^2} \sum^{L}_{n,m=-L}~\frac{1}{k^2_n + k^2_m + \eta^2}.
\end{equation}
The wave numbers are defined as 
\begin{equation}
k^{}_n = \frac{\pi n}{L},
\end{equation}
where $n\in{\mathds Z}$, which corresponds to the periodic boundary conditions. However, using periodic boundary conditions is by no means obvious nor obligatory. Other boundary conditions may well be employed, which would lead to the different scaling behavior (cf. Ref.~[\onlinecite{PhysRevB90}]). The relation between the dimensionless quantity $L$ and physical length $\ell$ is established via
\begin{equation}
\ell = \frac{\hbar v^{}_F}{E^{}_b}L,
\end{equation}
where $E^{}_b$ is the band width. In particular, for $E^{}_b=0.75$eV which we use below for fitting purposes we get 
\begin{equation}
\frac{\hbar v^{}_F}{E^{}_b} \sim 2.9 {\rm \AA{}},
\end{equation}
which is slightly larger than the carbon lattice spacing $\displaystyle a\sim2.4{\rm\AA{}}$. The disorder strength is measured in units of $(\hbar v^{}_F)^2$, therefore the quantity $s=(\hbar v^{}_F)^2$ represents the appropriate reference scale. 

\vspace{1mm}
\no
Then the main goal is to determine the $\beta$--function
\begin{equation}
\beta = \frac{d \log\bar\sigma}{d\log L}, 
\end{equation}
a function which describes the behavior of $\bar\sigma$ under a change of the linear system size $L$. 
Here we use a regularization scheme which fixes the maximum number of accounted modes to $(2L+1)^2$.
The continuous limit of the momentum integral is recovered for $L\to\infty$. If periodic boundary conditions are used, then the main contribution to the conductivity at small scales comes from the zero modes with $n=0$ and $m=0$. In this case the conductivity approaches the plateau of the infinite sample
from above. This scenario was recently realized experimentally in Ref.~[\onlinecite{Heer2014}].

\begin{figure*}[t]
\includegraphics[height=5.4cm]{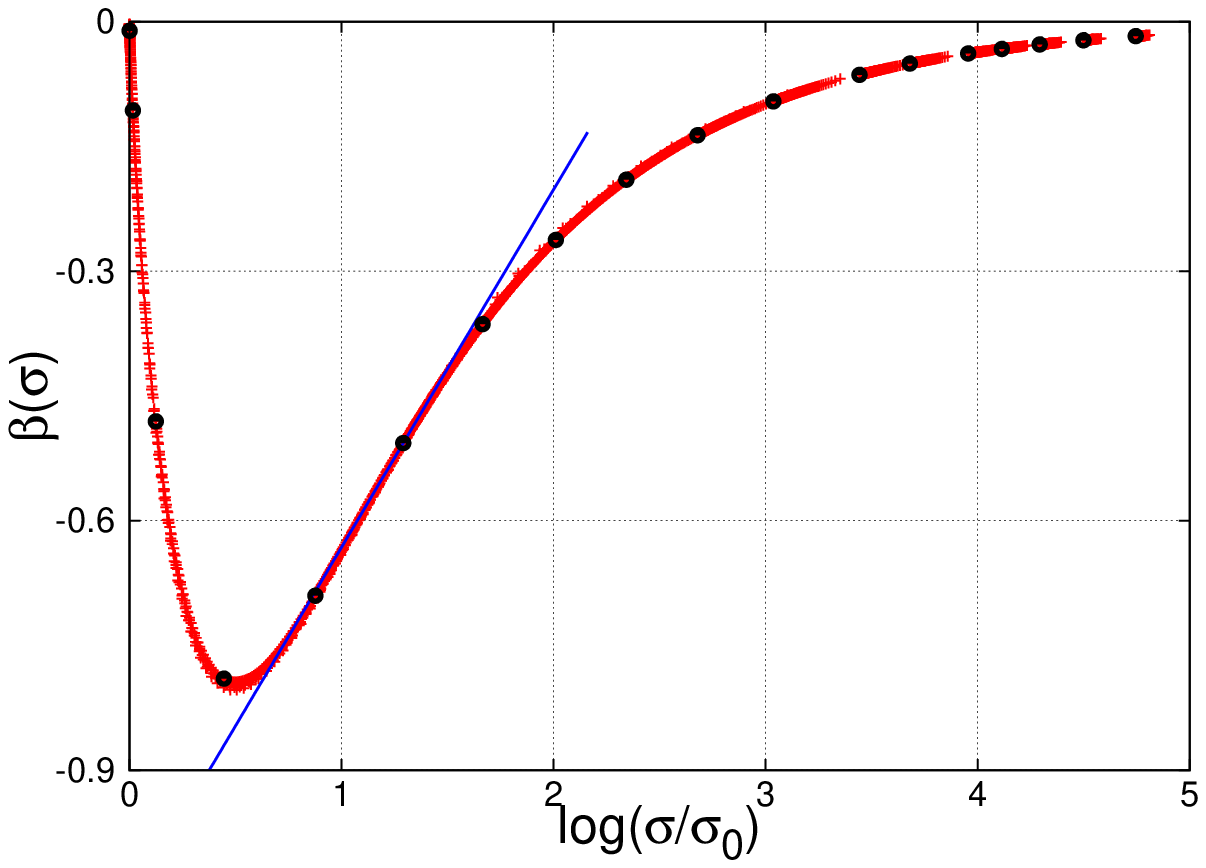}
\hspace{2mm}
\includegraphics[height=5.4cm]{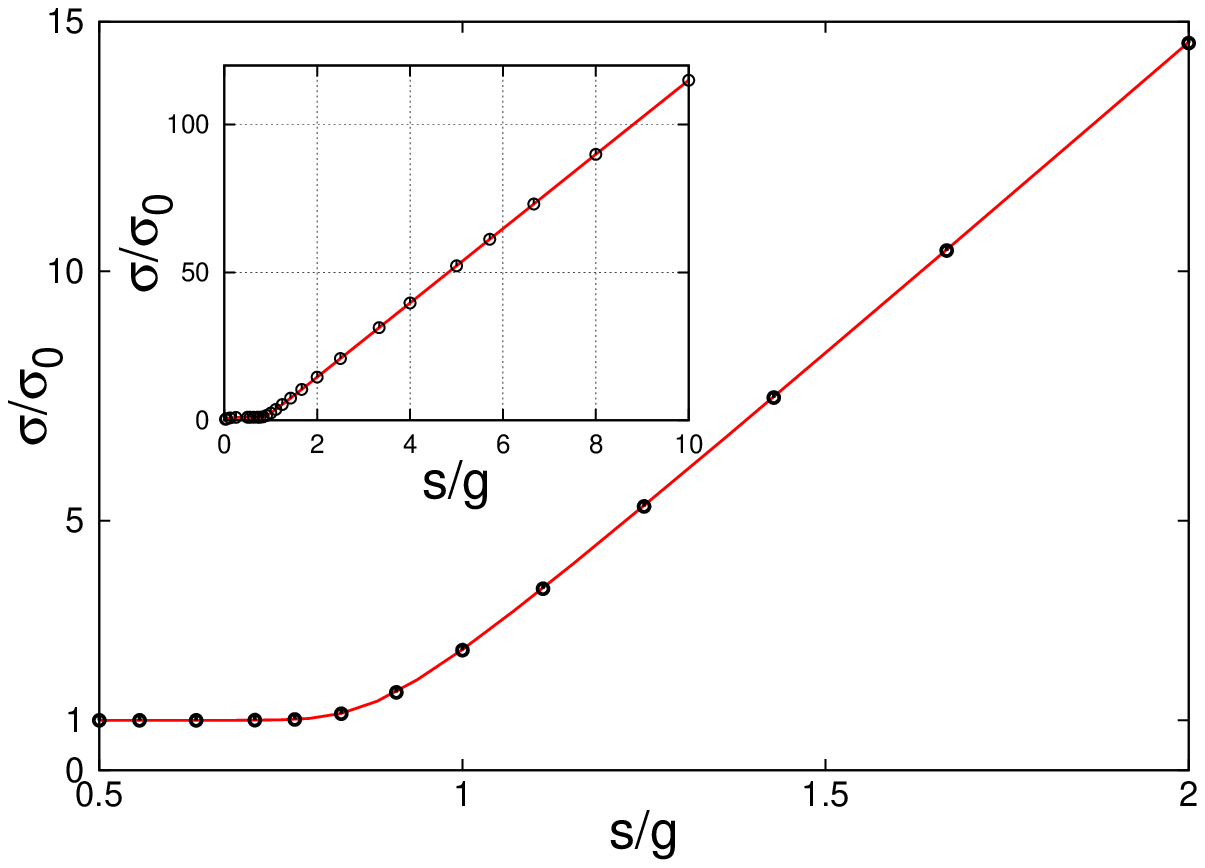}
\caption{(Color online) 
Left: $\beta$--function calculated on the square lattice with periodic boundary conditions. Each piece of the graph is calculated for a particular disorder strength with black dots giving the $\beta$--function at the maximal number of modes. The solid (blue) line emphasizes the area of  logarithmic decrease. 
Right: The conductivity dependence on inverse disorder strength extracted from the $\beta$--function. The conductivity is given in units of  $\sigma^{}_0 = 4/\pi e^2/h$. Main figure: Large $g$--asymptote with the distinct crossover region between the plateau and the regime with a linear growth. Black dots are the same from the figure on the right. Inset: Same data set depicted over the entire range of the disorder strength used for calculating the $\beta$--function. The linear behavior continues to the very weak disorder. 
}
\label{fig:Scaling}
\end{figure*}

\vspace{1mm}
\no
At first we study the case of a finite size sample which confines a 2D Dirac electron gas subject to a random potentials of different disorder strength. 
Choosing $L=120$ we solve Eqs.~(\ref{eq:DiffDiscret}) and (\ref{eq:SPDiscret}) for $g/s=0.1... 2$. This gives the scaling plot of the $\beta$--function 
depicted in Fig.~\ref{fig:Scaling}. Every piece of this graph corresponds to the particular disorder strength $g$. One recognizes a wide area where the
$\beta$--function decreases logarithmically and, therefore, obeys the one--parameter scaling behavior predicted by the famous scaling theory of Abrahams 
{\it et al.} in Ref.~[\onlinecite{Abrahams1979}]. For a sufficiently small conductivity, though, the flow turns up towards the fixed point which 
corresponds to the finite conductivity plateau. When extracted from the
scaling plot, the behavior of the conductivity as a function of $s/g$ exhibits two distinct regimes (cf. right--hand side of Fig.~\ref{fig:Scaling}): 
The linearly decreasing regime at disorder strength smaller than the hopping amplitude, i.e. $g<s$ and a plateau-like regime of nearly constant 
conductivity at disorder strengths above $s$.  Even for disorder as strong as $g=2s$, it still can be considered as a weakly disordered electron gas. 
However, the validity of our approach for larger values of $g$ is not guaranteed. Interestingly, the conductivity reveals a similar behavior if 
plotted versus the ratio $1/(\eta L)^2$. The effect of increasing $L$ reveals a broadening of the plateau region down to smaller values of $g$, 
while leaving the slope of the linear part at the same value. It is expected that in the limit $L\to\infty$ it must stretch over the 
whole range of disorder strengths, 
but even for the largest numerically accessible lattice sizes the crossover value is found at $g\sim0.8s$, i.e. it does not change significantly.

\begin{figure*}[t]
\includegraphics[height=5.4cm]{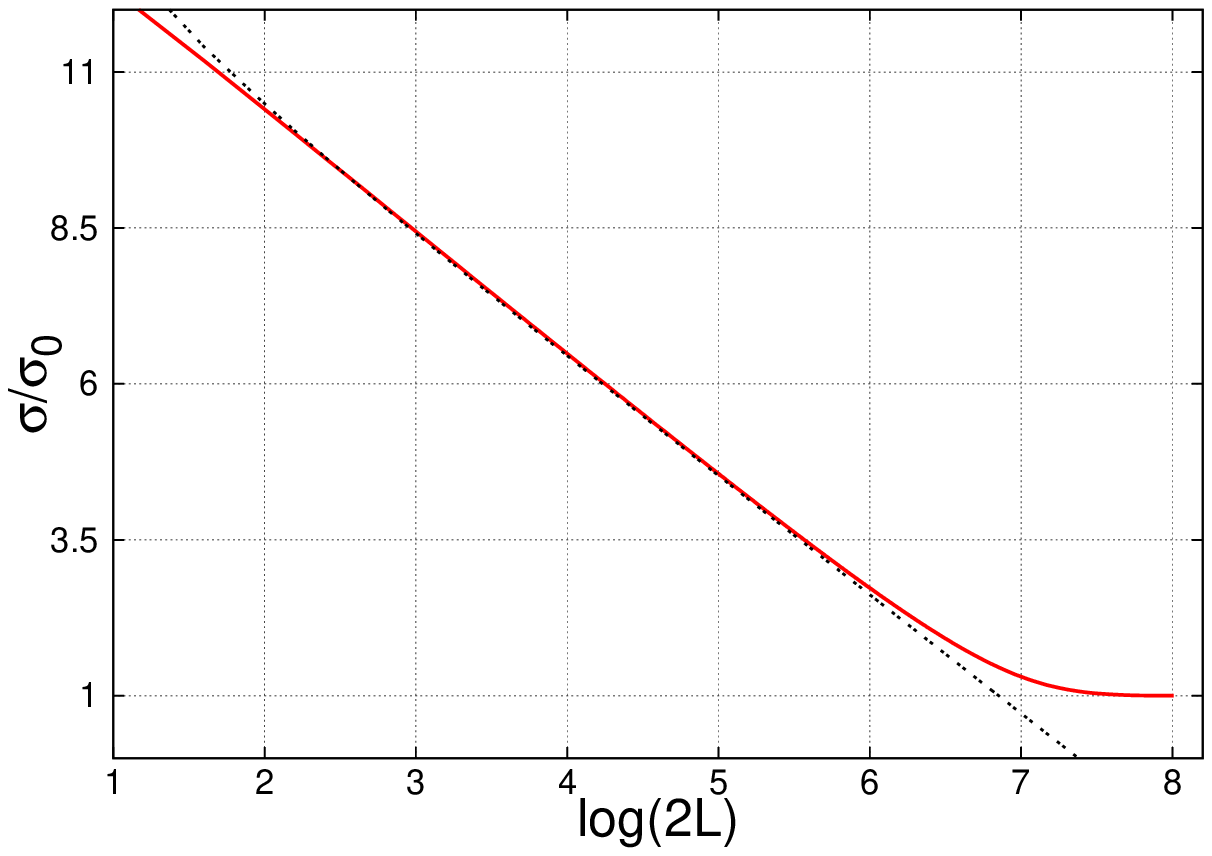}
\hspace{2mm}
\includegraphics[height=5.4cm]{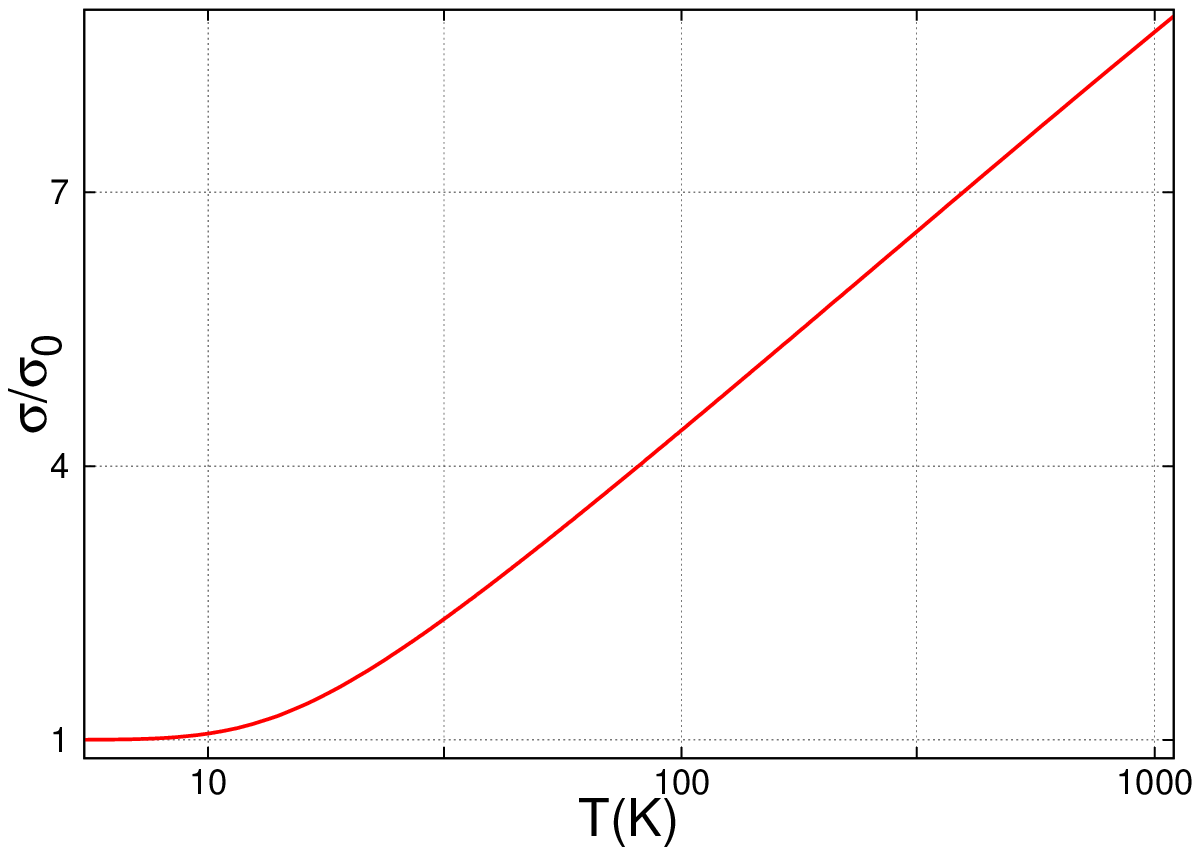}
\caption{(Color online) 
Left: Sample size dependence of the conductivity for the fixed value of the disorder strength $g=0.9s$. The conductivity reveals a broad region of logarithmic decrease which can be fitted with Eq.~(\ref{eq:LogFit}).
Right: Same data replotted as function of temperature using Eq.~(\ref{eq:Subst}). For the particular disorder strength, the crossover temperature which separates the plateau from the logarithmic region is found at roughly 15K. 
}
\label{fig:SizeTemp}
\end{figure*}

\vspace{1mm}
\no
In the second scenario the disorder strength is kept constant and the number of modes is gradually increased. For the case of weak disorder (e.g., for $g=0.3s$), 
the plateau cannot be reached for a moderate sample size, and the conductivity reveals the steady logarithmic decline, which is characteristic for the  weak
localization regime. For moderate disorder (e.g., for $g=0.9s$), the crossover into the plateau regime is reached roughly at $L\sim 10^3$, 
revealing a broad area of logarithmic decrease which can be fitted with the formula 
\begin{equation}
\label{eq:LogFit} 
\bar\sigma/\sigma^{}_0 \sim C - a\log(2L),
\end{equation}
where the slope $a\sim2$, specific for the case of orthogonal ensemble~[\onlinecite{Hikami1980}], is roughly the same for not too large values of $g$,  while the constant $C$ depends strongly on $g$, taking for instance $C\sim42$ for $g=0.3s$ and $C\sim12.7$ for $g=0.9s$. A temperature dependence of the conductivity
can be included as a finite cut-off for the sample size, using the substitution 
\begin{equation}
\label{eq:Subst}
 L = \frac{E^{}_b}{k^{}_B T}
 \ .
\end{equation}
This behavior can be seen on the right-hand side of Fig.~\ref{fig:SizeTemp}, where the band width is chosen to be $E^{}_b=0.75$eV. 
There is a crossover temperature of roughly 15K and an increase of the conductivity up to 4 times at room temperature. 
The overall shape of the temperature dependent conductivity, the value of the crossover temperature, 
as well as the prefactor of the logarithmic regime are in good agreement with recent experimental observations, performed on graphene with 
different degrees of disorder~[\onlinecite{Kogan2015,Fuhrer2009,Fuhrer2011,Savchenko2010,Camassel2011,Liu2014,Novoselov2012,Novoselov2011}].

\section{Discussion and conclusions}
\label{sec:Discussion}

\begin{figure*}[t]
\includegraphics[height=8.cm]{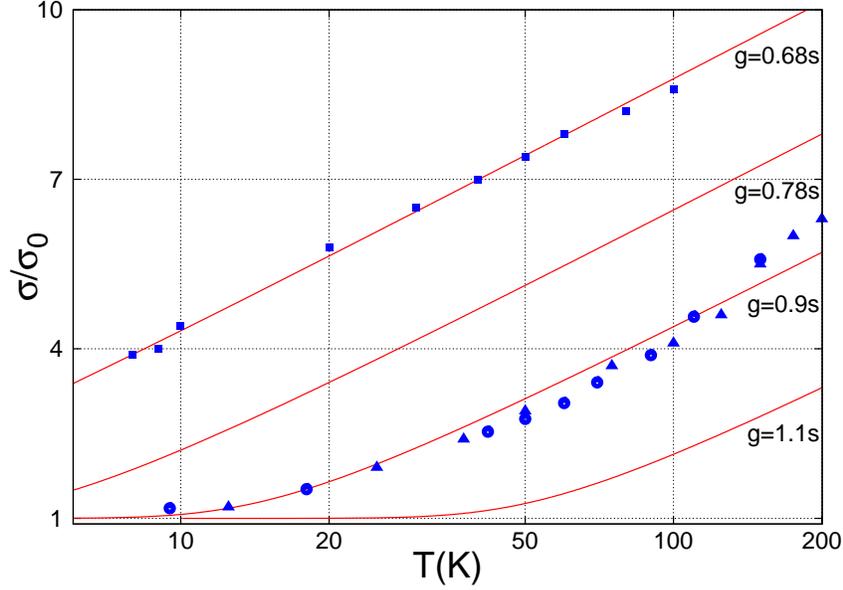}
\caption{(Color online) 
Fit of the experimental data with Eqs.~(\ref{eq:DiffDiscret}), (\ref{eq:SPDiscret}), and (\ref{eq:Subst}) evaluated with the maximal number of modes $L=1500$. The experimental data are extracted from Fig.~1(d) in Ref.~[\onlinecite{Liu2014}] (squares, the experimental data are shifted up by  $\sigma^{}_0$), the upper curve from Fig.~3(b) in Ref.~[\onlinecite{Novoselov2012}] (triangles, the experimental data are  sifted down by  0.9$\sigma^{}_0$), and second curve from Fig. 2(b) in Ref.~[\onlinecite{Novoselov2011}] (circles, where the experimental data are shifted up by  $0.5\sigma^{}_0$). The shifts are needed in order to match the experimental data to the charge neutrality, since they are published as raw data. For more details see the respective articles.}
\label{fig:Fit}
\end{figure*}

\no
A big challenge in the experimental investigation of transport in graphene and other realizations of a 2D Dirac electron gas is controlling
the amount of disorder. A possible way to create stronger disorder in the samples is by a bombardment with fast ions. We can use the results of such
experiments and compare them with our theoretical results. A considerable amount of published data can be fitted with the expressions in Eqs.~(\ref{eq:DiffDiscret}), (\ref{eq:SPDiscret}), and (\ref{eq:Subst}),
assuming a moderate disorder strength $g$. Fig.~\ref{fig:Fit} shows an example of such fits at low temperature. Here, the experimental data were 
extracted from Refs.~[\onlinecite{Liu2014}], [\onlinecite{Novoselov2012}], and [\onlinecite{Novoselov2011}], which appear as squares, triangles, and rings, respectively.

\vspace{1mm}
\no
The agreement between the experiment and our theory is good. Moreover, at first glance the fit of experimental data to our theoretical formulas requires two
parameters (i.e., disorder strength $g$ and band width $E^{}_b=k^{}_BTL$) (cf. Eq.~(\ref{eq:Subst})). However, these parameters cannot be measured independently, 
since they enter the scattering rate via
\begin{equation} 
\eta \sim E^{}_b ~ e^{-T/g}
\ ,
\end{equation}
which is directly accessible experimentally via the scattering time measurements: $\tau=\hbar/\eta$. From this point of view, our results represent in fact one--parameter fits. 

\vspace{1mm}
\no
The second group of reported experimental data (e.g., Ref.~[\onlinecite{Fuhrer2009,Kogan2015}] and some curves from [\onlinecite{Novoselov2012}])
cannot be fitted under assumption of weak or moderate disorder (i.e., for $g\leqslant2s$). In fact, they still can be fitted reliably well for 
much larger values of $g$ and much smaller bandwidths (i.e., $E^{}_b\ll t$) but the validity of the weak scattering assumption in this regime is 
not given anymore. 

\vspace{1mm}
\no
In conclusion, we have performed a thorough investigation of the DC conductivity of a weakly disordered 2D Dirac electron gas,
using perturbative ensemble averaging technique. Our results confirm that the temperature dependence of the conductivity of 
disordered 2D Dirac electron gases in finite size samples is characterized by two regimes in which it behaves distinctly different.
This is the regime of a logarithmic decay with sample size or decreasing temperature, observable at higher temperatures, and the regime 
of a nearly constant conductivity at lower temperatures. 
The plateau value as well as the crossover temperature between the two regimes are not universal but change from one sample to another, thus suggesting 
a strong dependence on the disorder, sample size and perhaps chemical doping. The obtained analytical expressions are very simple and do not reveal any
logarithmic divergences in the conductivity. Our results reflect correctly the shape and all features of the conductivity scaling behavior.

\section*{ACKNOWLEDGMENTS}

We acknowledge several discussions with E. Kogan.

\end{document}